\def\half{\frac{1}{2}}

\def\mb#1{\mbox{\boldmath{$#1$}}}
\def\eq#1{Eq.(\ref{#1})}

\documentclass[amsmath,amssymb,showkeys, showpacs]{revtex4}
\usepackage{graphicx}
\begin{document}

\hspace*{5 in}CUQM-152\\
\vspace*{0.4 in}
\title{Soft and hard confinement of a two-electron quantum system}

\author{Richard L. Hall$^1$, Nasser Saad$^2$, and K. D. Sen$^3$}
\address{$^1$ Department of Mathematics and Statistics, Concordia University,
1455 de Maisonneuve Boulevard West, Montr\'eal,
Qu\'ebec, Canada H3G 1M8}\email{richard.hall@concordia.ca}
\address{ $^2$ Department of Mathematics and Statistics,
University of Prince Edward Island, 550 University Avenue,
Charlottetown, PEI, Canada C1A 4P3.}\email{nsaad@upei.ca}
\address{$^3$ School of Chemistry, University  of Hyderabad 500046, India.}
\email{sensc@uohyd.ernet.in}

\begin{abstract}
\noindent A model physical problem  is studied in which  a system of two electrons is subject either to soft confinement by means
of attractive oscillator potentials or by entrapment within an impenetrable spherical box of finite radius $R.$
When hard confinement is present the oscillators can be switched off.
Exact analytical solutions are found for special parameter sets, and  highly accurate numerical solutions (18 decimal places) are obtained for general cases. Some interesting degeneracy questions are discussed at length. 
\end{abstract}
\maketitle
\section{Introduction}\label{intro}
\subsection{Confined atomic and molecular systems}
\noindent Quantum confined model systems incorporating different repulsive or attractive potential terms have  become the subject of  increasing research interest \cite{cruz,sen}. A number experiments involving, e.g. atoms and molecules under pressure, quantum dots, and atoms in metallofullerenes, are analysed using such models, which are essentially defined by the given shape and strength of the confining potential.  Further, the confined model potentials have been used to study the autoionization resonance states for the two-electron spherical quantum dots \cite{sajeev, genkin} and two electron atomic states \cite{chak}. Theoretical studies of confined systems generate valuable guidelines to the design and synthesis of new materials and future devices. It is therefore of interest to derive new exact solutions of the quantum model systems involving such confining potentials under certain special conditions.
\vskip0.1true in
\noindent We revisit a model by M. Taut \cite{taut} for a two-electron system in which the electrons are
softly confined by means of oscillator potentials.  We shall   consider the problem in $d\ge 2$ spatial dimensions
and find exact solutions for certain special choices of the parameters, and accurate numerical
solutions in other cases. We shall also study this system when it is confined to an impenetrable spherical box of radius $R$ in $\Re^d;$ we shall refer to this as `hard confinement', which persists even when the   oscillator interaction  is switched  off.  Thus we shall begin with only soft confinement in form of an attractive oscillator, and we go on later to implement hard confinement by applying to the same system a vanishing boundary condition at the surface of a finite sphere of radius $R.$ 
\vskip0.1true in
\noindent 
In order to place our work in perspective, we shall present here a brief
review on the confined $d$-dimensional H atom $V_c=-a/r$ and the isotropic harmonic
oscillator $V_h=b\,r^2$. The eigenspectrum of the spherically confined H atom
(SCHA) within impenetrable walls is characterized by novel degeneracy
\cite{pupyshev98} effects of three different kinds. Two of them are generated from the specific
choice of the radius of confinement $R$, chosen exactly at the
radial nodes corresponding to the free hydrogen atom (FHA) wave
functions. In the \emph{incidental degeneracy} case, the confined
$(\nu,\ell)$ state with the principal quantum number $\nu$ is
iso-energic with $(\nu+1,\ell)$ state of the FHA with energy
$-1/\{2(\nu+1)^2\}$ atomic units (a.u.), at an $R$ defined by the radial node in the
FHA. For example, the $(\nu,\ell)$ state corresponding to the
lowest energy value, when confined at the radius $R$ given by the
radial node the first excited \emph{free} state $(\nu+1,\ell)$,
increases in such a way that the confined-state energy becomes the same as
excited free-state energy. The specific node in question is given
by $R=0.24({2\ell}+d-1)({2\ell}+d+1)$. Such a degeneracy can be
realized at similar choices for $R$ where multiple nodes exist in
the second and higher excited states of a given $\ell$. However,
such closed analytical expressions for the radial nodes are not
available in the case of higher excited states. In the
\emph{simultaneous-degeneracy} case, on the other hand, for all
$\nu \ge \ell+2$, each  pair of confined states denoted by
$(\nu,\ell)$ and $(n+1,\ell+2)$ state, confined at the common
$R=0.24({2\ell}+d-1)({2\ell}+d+1)$, become degenerate. Note that
the pair of levels in the free state are nondegenerate. Both these
degeneracies have been shown \cite{pupyshev98} to result from the
Gauss relationship applied at a unique $R_c$ by the confluent
hypergeometric functions that describe the general solutions of
the SCHA problem. 
\vskip0.1true in
\noindent 
Finally, the interdimensional
degeneracy \cite{Herrick,HS} arises, as in the case of the free
hydrogen atom, due to the invariance of the Schr\"{o}dinger
equation under the transformation
$(\ell,d)~\rightarrow~(\ell\pm~1,~d\mp~2)$. In order to preserve
the number of nodes in the radial function, it is necessary simultaneously 
to make the transformation $\nu~\rightarrow~ \nu + 1$.
The \emph{incidental degeneracy} observed in the case of a
spherically confined isotropic harmonic oscillator (SCIHO) is
qualitatively similar to that of the SCHA. For example, the only
radial node in the first excited free state of any given $\ell$
for $d$-dimensional SCIHO is located at $R=\sqrt{(2\ell+d)/2}$. For
the multiple node states, the corresponding numerical values must
be used. However, the behavior of the two confined states at a
common radius of confinement is found to be interestingly
different \cite{senkd,spm}. In particular, for the SCIHO the pairs
of the confined states defined by $(\nu=\ell+1,\ell)$ and
$(\nu=\ell+2,\ell+2)$ at the common $R=\sqrt{(2\ell+d)/2}$ a.u.,
display for all $\nu$, a constant energy separation of
\emph{exactly} $2$ harmonic-oscillator units,~$2\hbar \omega$ ,
with the state of higher $\ell $ corresponding to the lower energy. It
is interesting to note that the two confined states at the
common $R$ with $\Delta \ell=2$, considered above
 contain different numbers of radial nodes.
 The condition for interdimensional
degeneracy\cite{Herrick,HS} due to the invariance of the
Schr\"{o}dinger equation remains the same as before. Recently, the
confined systems of the $d$-dimensional hydrogen atom \cite{Jaber}
and harmonic oscillator \cite{Ed} have been studied.
Problems involving short-range potentials in $d$ dimensions have recently been considered \cite{gusun,agboola}.
 In the light of the above discussion, it is interesting to study the various
aforementioned degeneracies in the free and spherically confined
$d$-dimensional potential generally given by $V(r) = V_c+V_h={a}/{r} +
b r^2$. 
\subsection{Formulation of the problem in $d$ dimensions}

We consider now the model potential of our interest in the present work. For a single particle the $d$-dimensional Schr\"odinger equation, in atomic units $\hbar=m=1$, with
a spherically symmetric potential $V(r)$ can be written as the following
\begin{equation}\label{eq1}
\left[-{1\over {2}}\,\Delta +V(r)\right]\psi(r)=E\psi(r),
\end{equation}
where $\Delta$ is the $d$-dimensional Laplacian operator and $r^2=\sum_{i=1}^d x_i^2$.
If the positions of two particles are $\mb{r}_1$ and $\mb{r}_2$, the Hamiltonian for the system in which the particles interact by means of a repulsive Coulomb term but are bound by oscillators is given by
\begin{equation}\label{eq2}
{\cal H} = -\half\Delta_{r_1}- \half\Delta_{r_2} + 2b^2r_1^2 + 2b^2r_2^2 + \frac{a}{|\mb{r}_1- \mb{r}_2|},\quad a,\,b >0,
\end{equation}
where $r_1 = |\mb{r}_1|$ and $r_2 = |\mb{r}_2|$.  By using a pair-distance coordinate $\mb{r} = \mb{r}_1- \mb{r}_2$  and a center-of-mass coordinate $\mb{R} = \half(\mb{r}_1+ \mb{r}_2)$, the Hamiltonian $H$ and potential $V$ for relative motion can be expressed in terms of $r = |\mb{r}|$ and $\Delta = \Delta_r,$ and may be written
\begin{equation}\label{eq3}
H = -\Delta + V(r), \quad V(r) = b^2 r^2 + \frac{a}{r}.
\end {equation}
In order to transform (\ref{eq1}) to the $d$-dimensional spherical coordinates $(r, \theta_1,\theta_2,\dots,\theta_{d-1})$, we first separate variables \cite{louck, atkin} using the representation
\begin{equation}\label{eq4}
\psi(r)=r^{-(d-1)/2}u(r)Y_{l_1,\dots,\,l_{d-1}}(\theta_1\dots\theta_{d-1}),
\end{equation}
where $Y_{l_1,\dots,\,l_{d-1}}(\theta_1\dots\theta_{d-1})$ is a normalized spherical harmonic with characteristic value $l(l+d-2)$, and $l=l_1=0,1,2,\dots$.  One then obtains the radial Schr\"odinger equation in the form
\begin{equation}\label{eq5}
(H-E)u(r) = \left[-\left({d^2\over dr^2}-{(k-1)(k-3)\over 4r^2}\right)+V(r)-E\right]u(r)=0,\quad\quad \int_0^\infty u^2(r)dr=1, u(0)=0,
\end{equation}
where $k=d+2l$.
\medskip

Since the spherical-harmonic factor $Y$ has definite parity \cite{atkin} given by $(-1)^{\ell},$ and the interaction in the present model does not depend on spin, the configuration-space wave function $\psi(r)$ can be multiplied by a two-particle spin factor with complementary permutation symmetry (singlet or triplet) so that the overall wave function for relative motion is antisymmetric under the exchange of the particle indices.   A center-of-mass factor $\phi(\mb{R})$ will not change this symmetry since it is in any case symmetric. Thus in the present model, all possible energy eigenstates must be considered for the two-body system in configuration space.
\medskip

We assume that the potential $V(r)$ is less singular than the centrifugal term so that
$$u(r)\sim A r^{(k-1)/2},\quad r\rightarrow 0,\quad \mbox{where $A$ is a constant}.$$
We note that the Hamiltonian and boundary conditions of (\ref{eq3}) are invariant under the transformation
$$(d,l)\rightarrow (d\mp2,l\pm 1).$$ Thus, given any solution for fixed $d$ and $l$, we can immediately generate others for different values of $d$ and $\ell.$ More particularly, the energy is unchanged if both  $k=2\ell+d$ and the number of nodes $n$ are held constant. Repeated application of this transformation produces a large collection of states, the only apparent limitation being a lack of interest in some values of $d$ (see, for example \cite{Doren}). In the present work, we consider the repulsive Coulomb plus a harmonic-oscillator potential
\begin{equation}\label{eq6}
V(r)={a\over r}+b^2r^2,\quad\quad b>0
\end{equation}
where $r=\|\mb{r}\|$, and the coefficients $a\ge 0$ and $b>0$ are both constant.
When we look at hard confinement inside a sphere of finite radius $R$, we can also consider removing the oscillator binding by choosing $b = 0$: in both cases of confinement, the spectrum of $H$ is entirely discrete.

\medskip

\subsection{Organization of the paper}
\noindent The present paper is organized as follows. In section 2, we discuss some general spectral features and bounds.  In section~3 we express the repulsive Coulomb
 potential term $a/r$ as the envelope of a family $\{\alpha(t)/r^2+\beta(t)\}_{t>0}$ of potential terms each of which admits an exact analytical solution. This application of envelope theory \cite{env1,env2,env3,env4,env5,env6} leads to a spectral upper bound formula  valid for all the soft-confined eigenvalues.  In section~4 we briefly review the asymptotic iteration method (AIM) for solving a second-order linear differential equation. We also discuss necessary and sufficient conditions for certain classes of differential equations with polynomial coefficients to have polynomial solutions. In sections 5 and 6, we use AIM to study how the
eigenvalues depend on the potential parameters $a$ and $b$. In each of these sections, the results obtained are of two types: exact analytic results that are valid when certain
parametric constraints are satisfied, and accurate numerical values for arbitrary sets of potential parameters. In section 5 there is only soft confinement effected by a non-zero oscillator term; in section 6 we consider `hard confinement', that is to say when the same system is confined to the interior of an impenetrable spherical box of radius $R.$
\section{Some general spectral features and analytical energy bounds}\label{bounds}
We shall show shortly that the Hamiltonian $H$ is bounded below. Indeed provided that $b>0$ or $R < \infty$, the entire spectrum is discrete. The eigenvalues of $H$ may therefore be characterized variationally.  We first discuss the open system where $R = \infty.$ The eigenvalues $E_{n,\ell}^d = E(a,b)$ are monotonic in each parameter $a$ and $b$ as a direct consequence of the monotonicity of the potential $V$ in these parameters. Indeed, since $\partial V/\partial a = 1/r >0$ and $\partial V/\partial b = 2br^2 > 0$, it follows from the Hellmann-Feynman theorem that
\begin{equation}\label{monotoneEab}
\frac{\partial E(a,b)}{\partial a} > 0\quad {\rm and}\quad\frac{\partial E(a,b)}{\partial b} > 0.
\end{equation}
Elementary scaling arguments can be used to reduce the dimension of the parametric space. For example we obtain
\begin{equation}\label{scale}
E(a,b) = \frac{1}{\sigma^2}E(\sigma a, \sigma^2b),\, \sigma >0,\quad{\rm thus}~E(a,b) = bE(a b^{-\half},\,1).
\end{equation}

\medskip
The generalized Heisenberg uncertainty relation  may be expressed  \cite{GS,RS2} for dimension $d\ge 3$  as the operator inequality $-\Delta > (d-2)^2/(4r^2).$ This allows us to construct the following lower energy bound
\begin{equation}\label{lbound}
E > {\mathcal E} = \min_{r>0}\left[\frac{(d-2)^2}{8r^2} + \frac{a}{r} + b^2 r^2\right].
\end{equation}
Provided $b > 0,$ this lower bound is finite for all $a$.  It also obeys the same scaling and  monotonicity laws as $E$ itself. But the bound is weak. For potentials such as $V(r)$ that satisfy $\frac{d}{dr}(r^2\frac{dV}{dr}) > 0,$  Common has shown \cite{common} for the ground state in $d=3$ dimensions that $\langle-\Delta\rangle > \langle 1/(2r^2)\rangle,$ but the resulting energy lower bound  is still weak. Lower and upper bounds on the ground-state energy $E$ are provided in $d$ dimensions, for example, by a Gaussian trial wave function $\phi(r)=c\,r^{\frac{d-1}{2}}e^{-\half\alpha r^2}$ used here in two distinct ways. By the local energy theorem \cite{thirring} we have (for $\alpha$ initially sufficiently small) we find the lower energy bound:
\begin{equation}\label{elet}
E \ge E_L = \max_{0<\alpha<\alpha_1}\,\min_{r>0}\,\left[\frac{H\phi}{\phi}\right] = \max_{0<\alpha<\alpha_1}\,\min_{r>0}\,\left[-\alpha^2 r^2 +\alpha d + \frac{a}{r} + b^2\,r^2\right].
\end{equation}
Meanwhile, if the constant $c$ is chosen so that normalization (apart from the angular terms), is provided by $\|\phi\|^2 = c^2\int_0^{\infty}\phi^2(r) dr = 1,$ then by the Rayleigh-Ritz theorem, we obtain the upper bound
\begin{equation}\label{err}
E \le E_U = \min_{\alpha}\left(\phi, H\phi\right) = \min_{\alpha} \left[ \frac{d\alpha}{2} + a\alpha^{\half}\,\frac{\Gamma(\frac{d-1}{2})}{\Gamma(\frac{n}{2})} + b^2\frac{d}{2\alpha} \right].
\end{equation}
 These bounds also bound the bottoms of the angular-momentum subspaces labelled by $\ell$: this is a consequence of the degeneracies between the states with the same value of $k = d+2\ell$, as we have mentioned above: practically speaking, if $\ell \ne 0,$ one can increase $d$ to $k$ and set $\ell=0$ in order to use the bound formulas (\ref{elet},\ref{err}) for the energies. For the $d$-dimensional oscillator $a=0$, we observe that the $d$-dimensional bounds coalesce to the exact energy $E = \min\limits_{\alpha}\,\half[d\alpha + b^2/\alpha]= db.$  A specific example is provided by the case $a=b=1$ and $d = 3$ for which we find
\begin{equation}\label{eul}
 E^L = 3.79049 < E = 4.057877 < 4.07988 = E^U,
\end{equation}
where the accurate estimate $E$ was found numerically, by shooting methods.

\medskip
More generally, we shall sometimes use the convention of atomic physics in which, even for non-Coulombic central potentials, a `principal quantum number' $\nu$ is used and defined
by
\begin{equation}\label{nu}
\nu = n+ \ell + (d-1)/2,
\end{equation}

Now we turn to the case of hard confinement $R <\infty$ and we write $E = E(a,b,R).$
The monotonicity with respect to the  box size $R$ may be  proved by a variational argument.
 Let us consider two box sizes, $R_1 < R_2$ and an angular momentum subspace labelled by a fixed $\ell.$
We extend the domains of the wave functions in the finite-dimensional subspace spanned by the first $N$ radial eigenfunctions for $R = R_1$ so that the new space $W$ may be used to study the case $R = R_2$. We do this by defining the extended eigenfunctions so that $\psi_i(r) = 0$  for $R_1 \le r\le R_2.$  We now look at $H$ in $W$ with box size $R_2$.  The minima of the energy matrix $[(\psi_i,H\psi_j)]$ are the exact eigenvalues for $R_1$ and, by the Rayleigh-Ritz principle, these values are one-by-one upper bounds to the eigenvalues for $R_2.$ Thus, by  formal argument we deduce what is perhaps intuitively clear, that the eigenvalues increase as $R$ is decreased, that is to say
\begin{equation}\label{montoneER}
\frac{\partial E(a,b,R)}{\partial R} < 0.
\end{equation}
If a very special box is now considered, whose size $R$ coincides with any radial node of the corresponding $R=\infty$ problem, then these two problems share an eigenvalue exactly. This is an example of a very general relation which exists between constrained and unconstrained eigensystems, and, indeed, also between two constrained systems with different box sizes.

\section{Upper-bound energy formula for soft confinement derived by the envelope method}
\begin{figure}[ht]
\centering
\includegraphics[scale=1.2]{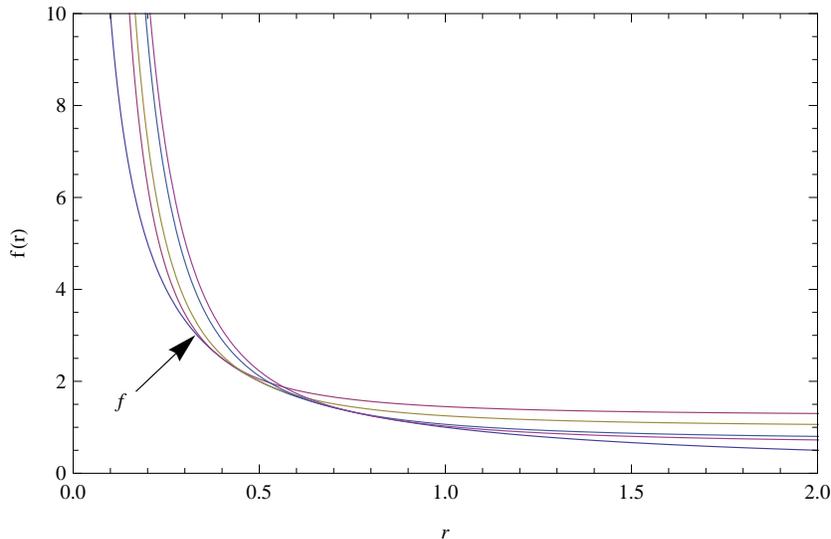}
\caption{The Coulomb term $f(r) = a/r$ in the potential is represented as
the envelope of an upper family of tangential potentials
 $\{f^{(t)}(r)=\alpha(t)/r^2 +\beta(t)\,\quad t>0\}$. When a tangential potential $f^{(t)}(r)$ is used in place of $f(r)$, Schr\"odinger's equation is exactly soluble. In the envelope method, the best choice is made by optimizing over the point of contact $t$.}\label{figenv}
\end{figure}

We first consider the radial eigenvalue problem for a generalized harmonic oscillator
we have
\begin{equation}\label{hoeq}
\left[-{d^2\over dr^2}+{\lambda(\lambda+1)\over r^2}+b^2r^2\right]u(r)=Eu(r),\quad\quad E = (4n +2\lambda+3)b,
\end{equation}
where $\int_0^\infty u^2(r)dr=1$ and $u(0)=0.$  Thus, for example, in $d\ge 2$ dimensions, we have
\[
\lambda = L = \ell + \frac{d-3}{2},\quad {\rm and}\quad E = (4n+ 2\ell +d)b,\quad n,l = 0,1,2,\dots
\]
Experience with the envelope method \cite{env1,env2,env3,env4,env5,env6} suggests that we express the repulsive Coulomb term $a/r$ as the envelope of a family of curves of the form $\{f^{(t)}(r) =\alpha(t)/r^2 + \beta(t)\}_{t>0},$ where $ r=t$ is the point of contact between the target $f(r) = a/r$ and a tangential curve $f^{(t)}(r).$  The idea behind this is to think of $a/r$ as a smooth  transformation $a/r = f(r) = g(1/r^2)$ of $1/r^2$ whose tangents $f^{(t)}(r)$ all lie above $a/r$ because
$g(X) = \sqrt{X}$ is concave; meanwhile the tangential potentials are Schr\"odinger--soluble. In order to find the coefficients $\alpha$ and $\beta$ we require at contact point $r=t$ that $f(r)$ and $f^{(t)}(r)$ agree in value and slope. Thus we have
\begin{equation}\label{tangents}
\frac{a}{t} = \frac{\alpha}{t^2}+\beta\quad {\rm and}\quad -\frac{a}{t^2} = -\frac{2\alpha}{t^3}\quad\quad
\Rightarrow\quad\quad \alpha = \frac{2t}{2} \quad {\rm and}\quad \beta = \frac{a}{2t}.
\end{equation}
Meanwhile, since $g$ is concave, we know that each tangential curve lies above $f(r)$, that is to say
\begin{equation}\label{envineq}
\frac{a}{r} \leq \frac{\alpha(t)}{r^2} +   \beta(t),\quad t >0.
\end{equation}
The geometrical situation is illustrated for the problem at hand in Fig.~\ref{figenv}.
In order to obtain the upper energy bound it remains to replace the Coulomb term in \eq{hoeq} by the right side of \eq{envineq} and solve the resulting equation
for the energy, shifted by $\beta.$  We first determine $\lambda$ from the equation
\[
\lambda(\lambda+1)=L(L+1) + \alpha\quad\quad \Rightarrow \quad \lambda = \left(\left(\ell +\frac{d-2}{2}\right)^2+\alpha\right)^{\half}-\half
\]
The final expression for this bound is as follows:
\begin{equation}\label{eu}
E \le E^{U} = \min_{t >0}{\mathcal E}^{U}(t),\quad {\mathcal E}^{U}(t)=
(4n+2)b + 2b\left[\left(\ell+\frac{d-2}{2}\right)^2 +\frac{at}{2}\right]^{\half} + \frac{ab}{2t}.
\end{equation}
The function ${\mathcal E}^{U}(t)$ is U-shaped with one minimum that is easy to find. What we have in effect is a continuous family of upper bounds, and the minimization in \eq{eu} selects the best of these for the given parameter set and eigenvalue.

For the ground state with $a = b = 1,$ we obtain the higher bound $E^U=4.2287$ than the $E^U = 4.07988 $ provided by a
scale-optimized Gaussian trial function in \eq{eul}, but of course the simple general formula \eq{eu} is valid for every eigenvalue.  By using AIM we shall find very accurate numerical eigenvalues for soft confinement in section~5, where some are tabulated in Table~\ref{table:difdn3}. We exhibit now two more upper bounds otained by \eq{eu}, along with accurate values in parentheses from Table~\ref{table:difdn3}: $E_{30}^{4}=E_{31}^{2}=(16.649791)<16.9444$; $E_{50}^{7} = E_{51}^{5} = E_{52}^{3} = (27.460231)<27.6228.$

\section{The asymptotic iteration method and some related results}\label{AIM}
\noindent The asymptotic iteration method (AIM) was originally
introduced \cite{aim} to investigatAe the solutions of differential
equations of the form
\begin{equation}\label{eq5}
y''=\lambda_0(r) y'+s_0(r) y,\quad\quad ({}^\prime={d\over dr})
\end{equation}
where $\lambda_0(r)$ and $s_0(r)$ are $C^{\infty}-$differentiable
functions. A key feature of this method is to note the invariant
structure of the right-hand side of (\ref{eq5}) under further
differentiation. Indeed, if we differentiate (\ref{eq5}) with
respect to $r$, we obtain
\begin{equation}\label{eq6}
y^{\prime\prime\prime}=\lambda_1 y^\prime+s_1 y
\end{equation}
where $\lambda_1= \lambda_0^\prime+s_0+\lambda_0^2$ and
$s_1=s_0^\prime+s_0\lambda_0.$ If we find the second derivative of
equation (\ref{eq5}), we obtain
\begin{equation}\label{eq7}
y^{(4)}=\lambda_2 y^\prime+s_2 y
\end{equation}
where $\lambda_2= \lambda_1^\prime+s_1+\lambda_0\lambda_1$ and
$s_2=s_1^\prime+s_0\lambda_1.$ Thus, for $(n+1)^{th}$ and
$(n+2)^{th}$ derivative of (\ref{eq5}), $n=1,2,\dots$, we have
\begin{equation}\label{eq8}
y^{(n+1)}=\lambda_{n-1}y^\prime+s_{n-1}y
\end{equation}
and
\begin{equation}\label{eq9}
y^{(n+2)}=\lambda_{n}y^\prime+s_{n}y
\end{equation}
respectively, where
\begin{equation}\label{eq10}
\lambda_{n}=
\lambda_{n-1}^\prime+s_{n-1}+\lambda_0\lambda_{n-1}\hbox{ ~~and~~
} s_{n}=s_{n-1}^\prime+s_0\lambda_{n-1}.
\end{equation}
From (\ref{eq8}) and (\ref{eq9}) we have
\begin{equation}\label{eq11}
\lambda_n y^{(n+1)}- \lambda_{n-1}y^{(n+2)} = \delta_ny {\rm
~~~where~~~}\delta_n=\lambda_n s_{n-1}-\lambda_{n-1}s_n.
\end{equation}
Clearly, from (\ref{eq11}) if $y$, the solution of
(\ref{eq5}), is a polynomial of degree $n$, then $\delta_n\equiv
0$. Further, if $\delta_n=0$, then $\delta_{n'}=0$ for all $n'\geq
n$. In an earlier paper \cite{aim} we proved the principal theorem
of AIM, namely
\vskip 0.1in

\noindent{\bf Theorem~1~\cite{aim}.} \emph{Given $\lambda_0$ and $s_0$ in
$C^{\infty}(a,b),$ the differential equation (\ref{eq5}) has the
general solution
\begin{equation}\label{eq12}
y(r)= \exp\left(-\int\limits^{r}{s_{n-1}(t)\over \lambda_{n-1}(t)} dt\right) \left[C_2
+C_1\int\limits^{r}\exp\left(\int\limits^{t}(\lambda_0(\tau) +
2{s_{n-1}\over \lambda_{n-1}}(\tau)) d\tau \right)dt\right]
\end{equation}
if for some $n>0$
\begin{equation}\label{eq13}
\delta_n=\lambda_n s_{n-1}-\lambda_{n-1}s_n=0.
\end{equation}
where $\lambda_n$ and $s_n$ are given by (\ref{eq10}).
}
\vskip0.1true in
\noindent Recently, it has been shown \cite{aim1} that the termination condition (\ref{eq13}) is necessary and sufficient for the differential
equation (\ref{eq5}) to have polynomial-type solutions of degree at most $n$, as we may conclude from Eq.(\ref{eq11}). Thus, using Theorem 1, we can now find the necessary and sufficient conditions \cite{saad} for the polynomial solutions of the differential equation
\begin{equation}\label{eq21}
(a_{3,0}r^3+a_{3,1}r^2+a_{3,2}r+a_{3,3})~y^{\prime \prime}+(a_{2,0}r^2+a_{2,1}r+a_{2,2})~y'-(\tau_{1,0} r+\tau_{1,1})~y=0,
\end{equation}
where $a_{k,j}, k=3,2,1, j=0,1,2,3$ are constants. These conditions have been reported in the following theorem:
\vskip0.1true in
\noindent{\bf Theorem 2~[\cite{saad} Theorem 5].} \emph{
The second-order linear differential equation (\ref{eq21}) has a polynomial solution of degree $n$ if
\begin{equation}\label{eq22}
\tau_{1,0}=n(n-1)~a_{3,0}+n~a_{2,0},\quad n=0,1,2,\dots,
\end{equation}
provided $a_{3,0}^2+a_{2,0}^2\neq 0$ along with the vanishing of $(n+1)\times(n+1)$-determinant $\Delta_{n+1}$ given by
\begin{center}
$\Delta_{n+1}$~~=~~\begin{tabular}{|lllllll|}
 $\beta_0~~$ & $\alpha_1$ &$\eta_1$&~& ~&~ &~\\
  $\gamma_1$ & $\beta_1$ &  $\alpha_2$&$\eta_2$&~&~&~ \\
~ & $\gamma_2$  & $\beta_2$&$\alpha_3$&$\eta_3$&~&~\\
$~$&~&$\ddots$&$\ddots$&$\ddots$&$\ddots$&~\\
~&~&~&$\gamma_{n-2}$&$\beta_{n-2}$&$\alpha_{n-1}$&$\eta_{n-1}$\\
~&~&~&&~$\gamma_{n-1}$&$\beta_{n-1}$&$\alpha_n$\\
~&~&~&~&$~$&$\gamma_{n}$&$\beta_n$\\
\end{tabular}
\end{center}
where all the other entires are zeros and
\begin{align}\label{eq23}
\beta_n&=\tau_{1,1}-n((n-1)a_{3,1}+a_{2,1})\notag\\
\alpha_n&=-n((n-1)a_{3,2}+a_{2,2})\notag\\
\gamma_n&=\tau_{1,0}-(n-1)((n-2)a_{3,0}+a_{2,0})\notag\\
\eta_n&=-n(n+1)a_{3,3}.
\end{align}
Here $\tau_{1,0}$ is fixed for a given $n$ in the determinant $\Delta_{n+1}=0$ (the degree of the polynomial solution). The coefficients of the polynomial solution $y_n(r)=\sum_{i=0}^n c_i r^i$ satisfy the four-term recursive relation
\begin{align}\label{eq24}
&(i+2)(i+1)a_{3,3}c_{i+2}+\left[i(i+1)a_{3,2}+(i+1)a_{2,2}\right]c_{i+1}+\left[i(i-1)a_{3,1}+ia_{2,1}-\tau_{1,1}\right]c_i\notag\\
&+ \left[(i-1)(i-2)a_{3,0}+(i-1)a_{2,0}-\tau_{1,0}\right]c_{i-1}=0.
\end{align}
}

\noindent In the next sections, we shall apply the result of theorem~2 to study the possible quasi-exact analytic solutions for the $d$-dimension Schr\"odinger equation (\ref{eq3}) for unconstrained and constrained Coulomb plus harmonic oscillator potential (\ref{eq4}). We shall also apply AIM, theorem~1, to obtain \emph{accurate} approximations for arbitrary potential parameters, again, for the unconstrained and constrained $d$-dimensional Schr\"odinger equation (\ref{eq3}).
\section{Exact and approximate solutions for soft confinement}\label{spec}
\subsection{Exact bound-state solutions for a repulsive Coulomb plus
harmonic oscillator potential in $d$-dimensions}

\noindent In this section, we consider the $d$-dimensional Schr\"odinger equation
\begin{equation}\label{eq25}
\left[-{d^2\over dr^2}+{(k-1)(k-3)\over 4r^2}+{a\over r}+b^2r^2\right]u_{nl}^d(r)=E_{nl}^d u_{nl}^d(r),\quad 0<r<\infty.
\end{equation}
In order to  solve this equation by means of  AIM, the first step is to
transform (\ref{eq25}) into the standard form (\ref{eq5}). To this end, we note that the differential equation (\ref{eq25}) has one regular singular
point at $r = 0$ and an irregular singular point at $r = \infty$.  Since for large $r$, the harmonic oscillator term dominates, the asymptotic solution of (\ref{eq25}) as $r\rightarrow \infty$ is $u_{r\rightarrow \infty}\sim \exp({-{b~r^2}})$; meanwhile the indicial equation of (\ref{eq25}) at the regular singular point $r=0$ yields
\begin{equation}\label{eq26}
s(s-1)-{1\over 4}(k-1)(k-3)=0,
\end{equation}
which is solved by
\begin{align*}
s_1={1\over 2}(3-k),\quad s_2={1\over 2}(k-1).
\end{align*}
The value of $s$, in Eq.(\ref{eq26}), determines the behavior of $u_{nl}^d(r)$ for $r\rightarrow 0$, and only $s>1/2$ is acceptable, since only in this case is the mean value of the kinetic energy finite \cite{landau}.
Thus, the exact solution of (\ref{eq25}) may assume the form
\begin{equation}\label{eq27}
u_{nl}^d(r)=r^{(k-1)/2}\exp(-b\,r^2/2)~f_n(r),\quad k=d+3l,
\end{equation}
where we note that $u_{nl}^d(r)\sim r^{(k-1)/2}$ as $r\rightarrow 0$. On substituting this ansatz wave function into (\ref{eq25}), we obtain the differential equation for $f_n(r)$ as
\begin{equation}\label{eq28}
r\,f_n''(r)+\left(-2\, b\, r^2+k-1\right)f_n'(r)-\left((k\,b-E_n)\,r+a\right)f_n(r)=0.
\end{equation}
This equation has a regular singular point at $r=0$ with exponents $\{0,2-k\}$ and an irregular point at infinity.
We may note first that  is a special case of the differential equation (\ref{eq21}) with $a_{3,0}=a_{3,1}=a_{3,3}=a_{2,1}=0$, $a_{3,2}=1$, $a_{2,0}=-2\,b,~a_{2,2}=k-1$, $\tau_{1,0}=k\,b-E_n$ and $\tau_{1,1}=a$. Thus, the necessary condition for the polynomial solutions of Eq.(\ref{eq28}) is
\begin{equation}\label{eq29}
E_{nl}^d=b\,(2n'+k),\quad n'=0,1,2,\dots.
\end{equation}
and the sufficient condition follows from the vanishing of the tridiagonal determinant $\Delta_{n+1}=0$, $n=0,1,2,\dots$, namely
\begin{center}
$\Delta_{n+1}$~~=~~\begin{tabular}{|lllllll|}
 $\beta_0~~$ & $\alpha_1$ &~&~& ~&~ &~\\
  $\gamma_1$ & $\beta_1$ &  $\alpha_2$&~&~&~&~ \\
~ & $\gamma_2$  & $\beta_2$&$\alpha_3$&~&~&~\\
$~$&~&$\ddots$&$\ddots$&$\ddots$&~&~\\
~&~&~&$\gamma_{n-2}$&$\beta_{n-2}$&$\alpha_{n-1}$&~\\
~&~&~&&~$\gamma_{n-1}$&$\beta_{n-1}$&$\alpha_n$\\
~&~&~&~&$~$&$\gamma_{n}$&$\beta_n$\\
\end{tabular}~~=~~0
\end{center}
where its entries are expressed in terms of the parameters of Eq.(\ref{eq28}) by
\begin{align}\label{eq30}
\beta_n&=a,\quad \alpha_n=-n(n+k-2),\quad \gamma_n=2b(n-n'-1), \quad \eta_n=0,
\end{align}
where $n'=n$ is fixed by the size of the determinant $\Delta_{n+1}=0$ and represent the degree of the polynomial solution of Eq.(\ref{eq28}). We may note that, since the off-diagonal entries $\alpha_i$ and $\gamma_i$ of the tridiagonal determinant satisfy the identity
$$\alpha_i\,\gamma_i>0,\quad \forall~~ i=1,2,\dots,$$
the latent roots of the determinant $\Delta_{n+1}$ are all real and distinct \cite{arscott}.  Further, we can easily show that the determinant defined by the entries  of (\ref{eq30}) satisfies a three-term recurrence relation
\begin{equation}\label{eq31}
\Delta_i=\beta_{i-1}\Delta_{i-1}-\gamma_{i-1}\alpha_{i-1}\Delta_{i-2}, \quad \Delta_0=1,~~\Delta_{-1}=0,\quad i=1,2,\dots
\end{equation}
which can be used to compute the determinant $\Delta_i$ (and thus the sufficient conditions) recursively, in terms of lower order determinants. In this case, however, we must fix $n'$ for each of the sub-determinants used in computing (\ref{eq31}). For example, in the case of $n'=n=1$ (corresponding to a polynomial solution of degree one),
we have
\begin{center}
$\Delta_{2}$~~=~~\begin{tabular}{|lr|}
 $a$ & $1-k$ \\
  $-2b$ & $a$ \\
\end{tabular}~~=$a^2-2b(k-1),$
\end{center}
meanwhile, $$\Delta_{2}=~\beta_{1}\,\Delta_{1}-\gamma_{1}\,\alpha_{1}\,\Delta_{0}=a(a)-(-2b)(-(k-1))(1)=a^2-2b(k-1),$$
that is, the condition on the potential parameters reads
\begin{equation}\label{eq32}
a^2-2b(k-1)=0.
\end{equation}
For $n'=2$ (corresponding to a second-degree polynomial solution of \eqref{eq28}), we have
\begin{center}
$\Delta_{3}$~~=~~\begin{tabular}{|lcr|}
 $a~~$ & $1-k$ &$0$\\
  $-4b$ & $a$ &  $-2k$\\
0& $-2b$  & $a$
\end{tabular}~=$\beta_{2}\,\Delta_{2}-\gamma_{2}\,\alpha_{2}\,\Delta_{1}$=$a\begin{tabular}{|lr|}
 $a~~$ & $1-k$ \\
  $-4b$ & $a$
\end{tabular}-(-2b)(-2k)(a)=a(a^2-4\,b\,(2k-1))$.
\end{center}
Consequently, we must have
\begin{equation}\label{eq33}
a(a^2-4\,b\,(2k-1))=0.
\end{equation}
In Table I, we give the conditions on the potential parameters to yield polynomial solutions, from theorem 2.
\vskip0.true in
\begin{table}[!h] \caption{Conditions on the parameters $a$ and $b$ for the exact solutions of Eq.(\ref{eq28}) with $E_{nl}^d=b\,(2n+k)$, $k=d+2l$. } 
\centering 
\begin{tabular}{l l l} 
\hline\hline 
$n$ & ~ &$\Delta_{n+1}=0 $\\ [0.5ex] 
\hline\hline 
0   &~& $a=0$\\ 
\hline 
1   &~& $a^2-2b(k-1)=0$\\ 
\hline
2  &~& $a(a^2-4\,b\,(2k-1))=0$\\ 
\hline
3   &~& $a^4 - 20 a^2 b k + 36 b^2 (k^2-1) =0$\\ 
\hline
4   & ~&$a (a^4 - 20\, a^2\,b\, ( 2 k+1) + 32\, b^2 (8 k^2+8k-7))=0$\\ 
\hline
5   & ~&$a^6 - 70 a^4 b (k+1) +
 4 a^2 b^2 (259 k (k + 2)-65)- 1800 b^3 (k-1) (k+1) (k+3) =0$\\ 
\hline
\hline 
\end{tabular}
\label{table:nonlin}
\end{table}

\noindent It must be clear that although $n$, the degree of the polynomial solution, it is not necessarily an indication as to the number of the zeros of the wave function (node number): further analysis of the roots of $f_n(r)$ is usually needed to compute the zeros of the wavefunction.
\vskip0.1true in
\noindent The polynomial solutions $f_{n'}(r)=\sum_{i=0}^{n'} c_ir^i$ can be easily constructed for each $n'$ since, in this case, the coefficients $c_i$ satisfy the three-term recurrence relation (see Eq.(\ref{eq31}))
\begin{align}\label{eq34}
 c_{-1}=0,\quad c_0=1,\quad c_{i+1}&={ac_i+2b\,(i-n'-1)c_{i-1}\over (i+1)(i+k-1)},~~ i=0,1,\dots,n'-1,
\end{align}
where $n'$ is the degree of the polynomial solution of the differential equation \eqref{eq31}.
The zeros of these polynomials characterize the nodes (vanishing
points) of the physical wavefunctions. When $n'=0$, $f_0(r)=1$.
For  $n'=1,i=0$, we have
$$c_1={a\over k-1},
$$
that is, for $r>0$,
\begin{equation}\label{eq35}
f_1(r)=1+{a\over k-1}\,r ,\quad \mbox{subject to}\quad a=\sqrt{2b(k-1)}>0.
\end{equation}
Thus, there is no root of $f_1(r)$ in $(0,\infty)$ and  the (normalized) wave function then reads
\begin{equation}\label{eq36}
u_{0l}^d(r)=\left[\dfrac{\sqrt{2k-2}\,\Gamma\left(\dfrac{k-1}{2}\right)+\dfrac{(2k-1)}{(k-1)}\Gamma\left(\dfrac{k}{2}\right)}{2b^{k/2}}\right]^{-\dfrac12} r^{\dfrac{k-1}{2}}\exp\left(-{b\,r^2\over 2}\right)~\left(1+\sqrt{\dfrac{2b}{k-1}}\,r\right),
\end{equation}
 represents the normalized ground-state solution of Schr\"odinger's equation supported by the potential
\begin{equation}\label{eq37}
V_0(r;\sqrt{2b(k-1)},b)=\dfrac{\sqrt{2b(k-1)}}{r}+b^2\, r^2,
\end{equation}
with ground-state energy
\begin{equation}\label{eq38}
E_{0l}^d\equiv E_{0l\pm 1}^{d\mp 2}={b(2+k)}.
\end{equation}

\noindent For second-degree polynomial solution, $n'=2,i=0,1$, we have for the polynomial solution, $f_2(r)=c_0+c_1r+c_2r^2$, the coefficients
\begin{align*}
c_0=1,\quad c_1&={a\over k-1}\quad \mbox{and}\quad c_2={a^2+4b(1-k)\over 2k(k-1)},\qquad\mbox{subject to}\qquad a=2\sqrt{b(2k-1)}>0,
\end{align*}
and the polynomial solution then reads
\begin{equation}\label{eq39}
f_2(r)=\dfrac{1}{k-1}\left(2\,b\,r^2+2\sqrt{b\,(2k-1)}\,r+k-1\right).
\end{equation}
It is easily demonstrated that the (normalized) wave-function of the lowest energy state takes the form
\begin{equation}\label{eq40}
u_{0l}^d(r)=\dfrac{\sqrt{2b^{k/2}}\,\,r^{(k-1)/2}\,\,e^{-{b\,r^2}/{2}}}{\sqrt{(1-4k+8k^2)\Gamma(k/2)+8k\sqrt{2k-1}\Gamma((k+1)/2)}}~\left(2\,b\,r^2+2\sqrt{b\,(2k-1)}\,r+k-1\right),\quad b>0,
\end{equation}
and represents the ground state solution of Schr\"odinger's equation supported by the potential.
 \begin{equation}\label{eq41}
V_0(r;2\sqrt{b(2k-1)},b)=\dfrac{2\sqrt{b(2k-1)}}{r}+b^2\, r^2,\qquad 0<r<\infty
\end{equation}
The eigenvalue is given by
\begin{equation}\label{eq42}
E_{0l}^d\equiv E_{0l\pm 1}^{d\mp 2}=b(4+k).
\end{equation}
For a third-degree polynomial solution of equation \eqref{eq28}, $n'=3,i=0,1,2$, we have for the polynomial coefficients $f_3(r)=c_0+c_1r+c_2r^2+c_3r^3$ that
\begin{align*}
c_0=1,\quad c_1&={a\over k-1},\quad c_2={a^2+6b(1-k)\over 2k(k-1)},\quad c_3={a(a^2+2b(3-7k))\over 6(k-1)k(k+1)},\quad\mbox{subject to}\quad a=\sqrt{10kb\pm 2b\sqrt{16k^2+9}}.
\end{align*}
From this,  two polynomial solutions follow:
\begin{align}\label{eq43}
f_3^{+}(r)&=
1+\dfrac{\sqrt{10 b k+2b\sqrt{9+16 k^2}}}{k-1}\, r+\dfrac{b (3+2 k+\sqrt{9+16 k^2} )}{k(k-1) }\, r^2+\frac{b\left(3-2 k+\sqrt{9+16 k^2}\right) \sqrt{10 b k+2b\sqrt{9+16 k^2}} }{3 k \left(k^2-1\right)}\, r^3
\end{align}
and
\begin{align}\label{eq44}
f_3^{-}(r)&=
1+\dfrac{\sqrt{10 b k-2b\sqrt{9+16 k^2}}}{k-1}\, r+\dfrac{b (3+2 k-\sqrt{9+16 k^2} )}{k(k-1) }\, r^2-\frac{b\left(2 k-3+\sqrt{9+16 k^2}\right) \sqrt{10 b k+2b\sqrt{9+16 k^2}} }{3 k \left(k^2-1\right)}\, r^3
\end{align}
We may note here that the wave function
\begin{align}\label{eq45}
u_{0l}^{+\,d}(r)&=r^{\dfrac{k-1}{2}}e^{-\dfrac{b\,r^2}{2}}~f_3^+(r),
\end{align}
 has no real roots and thus represent the ground-state solution of  Schr\"odinger's equation supported by the potential
\begin{equation}\label{eq46}
V_0(r;\sqrt{10kb+2b\sqrt{16k^2+9}},b)=\dfrac{\sqrt{10kb+ 2b\sqrt{16k^2+9}}}{r}+b^2\, r^2,
\end{equation}
with corresponding eigenvalue given by
\begin{equation}\label{eq47}
E_{0l}^{d\pm}\equiv E_{0l\pm 1}^{d\mp 2}=b(6+k).
\end{equation}
However, the wave function
\begin{align}\label{eq48}
u_{1l}^{-\,d}(r)&=r^{\dfrac{k-1}{2}}e^{-\dfrac{b\,r^2}{2}}~f_3^-(r),
\end{align}
has one real root $r^*\in (0,\infty)$ computed by the zeros of $$f_3^-(r^*)=0,\qquad r^*>0.$$
In either case,  the normalization constant can be easily evaluated by using the relation
$$C_\pm^{-2}=\int_0^\infty \left[u_{nl}^{\pm d}(r)\right]^2dr\,\, \mbox{using the identity}\quad \int_0^\infty r^{\nu-1}\exp\left(-\mu r^p\right)dr=\dfrac{1}{p}\nu^{-\nu/p}\Gamma\left(\dfrac{\nu}{p}\right),\,\, \nu,p,\mu>0.$$
Similarly, using the recurrence relation, Eq.(\ref{eq31}), it is straightforward to compute explicitly the polynomial solutions of for higher order.
\vskip0.1true in
\noindent It is also interesting to note that equation \eqref{eq28} is a special case of the Biconfluent Heun (BCH) differential equation defined by the canonical form
\begin{equation}\label{eq49}
rf^{\prime\prime}(r) +(1+\alpha-\beta r-2r^2)f^\prime(r)+\left[(\gamma-\alpha-2)\,r-\dfrac12(\delta+(1+\alpha)\beta)\right]f(r)=0.
\end{equation}
A simple comparison between equations \eqref{eq28} and \eqref{eq47} allow to define the parameters, through a simple substitution  $z=\sqrt{b}\, r$, by
\begin{equation}\label{eq50}
\alpha=k-2,\quad \beta=0,\quad \gamma=\dfrac{E}{b},\qquad \delta = \dfrac{2a}{\sqrt{b}}.
\end{equation}
Thus, a solution of equation \eqref{eq28}, in terms of the Biconfluent Heun functions, is given by
\begin{equation}\label{eq51}
f(r)=H_B\left(k-2,0,\dfrac{E}{b},\dfrac{2a}{\sqrt{b}},\sqrt{b}\, r\right).
\end{equation}
The polynomial solutions of the equation \eqref{eq47} have been analysed by several authors \cite{hautot1969,rovder}. Their results by means of the conditions
  \begin{equation}\label{eq52}
\gamma-\alpha-2=2n',\quad n'=0,1,2,\dots \Longrightarrow E_{n'}=b(2n'+k),
\end{equation}confirm our polynomial solutions of equation \eqref{eq28}, and
\begin{equation}\label{eq53}
\Delta_{n'+1}=0,
\end{equation}
where $\Delta_{n'+1}=0$ is a polynomial in $a$ of degree $i=n'+1$ defined by
\begin{equation}\label{eq54}
\Delta_{i}-a\,\Delta_{i-1}+2b(i-1)(i+k-3)(i-n'-2)\Delta_{i-2}=0,\quad i\ge 1,\quad \Delta_{-1}=0,\quad \Delta_0=1.
\end{equation}
The roots of these polynomials gives the condition(s) on the potential's parameter $a$ as a function of $b$ and $k$. Having these values at our disposal, we evaluate the roots $R$ of the polynomial solutions
\begin{equation}\label{eq55}
 H_B\left(k-2,0,2n'+k,\dfrac{2a}{\sqrt{b}},\sqrt{b}\,  R\right)=0,\qquad k=d+2l
\end{equation}
to determines the number of nodes in the solution of the Schr\"{o}dinger equation \eqref{eq25}. This procedure can also used  to analyze the exact solutions of the confined potential $V(r)= a/r+b^2 r^2$ with an impenetrable spherical cavity at $r=R$. In Table \ref{table:nonlin2}, we report the value(s) of the parameter $a$ obtained using the determinant $\Delta_{n'+1}=0$, or roots of \eqref{eq55}, for the exact solutions of  the potential $V(r)=a/r+r^2$ (i.e. $b=1$) along with the number of roots (thus exhibiting the node number)  of the wave function solutions.

\begin{table}[!h] \caption{The value(s) of the parameter $a$ found from $\Delta_{n'+1}=0$ for the potential $V(r)=a/r+r^2$ and the roots of the polynomial solutions. } 
\centering 
\begin{tabular}{|l|l|r|r|} 
\hline\hline 
$n'$~($\Delta_{n'+1}=0$)  & $a$ &Roots & Type of solution \\ [0.5ex] 
\hline\hline 
0   & $a=0$ & N/A & Ground-state \\ 
\hline
1   & $a=2$ & N/A & Ground-state \\
\hline
2   & $a=2\sqrt{5}$ & N/A & Ground-state \\
\hline
3   & $a=\sqrt{30+6\sqrt{17}}$ & N/A & Ground-state \\
~& $a=\sqrt{30-6\sqrt{17}}$ & $R=1.447~082~228~754~501~502~2$& First-excited state\\
\hline
4   & $a=\sqrt{70+6\sqrt{57}}$ & N/A & Ground-state \\
~& $a=\sqrt{70-6\sqrt{57}}$ & $R=1.653~264~540~801~602~796~4$& First-excited state\\
\hline
5 & $a=14.450~001~026~965~667~202$ & N/A&Ground-state\\
~& $a=~8.050~661~272~517~918~496~6$ & $R_{}= 1.840~998~133~456~948~787~3$ & First-excited state\\
~ & $a=~2.526~721~867~533~372~270~5$ & $R_1= 1.146~288~753~895~025~008~6$&~\\ ~&~& $R_2=2.216~251~221~016~773~736~3$ &Second-excited state.\\
\hline 
\end{tabular}
\label{table:nonlin2}
\end{table}

\vskip0.1true in
\noindent For arbitrary values of the potential parameters $a$ and $b$,  we apply, in the next section, the asymptotic iteration method to solve Schr\"odinger's equation \eqref{eq25}.  First, we close this section by discussing the special case of the case $a=0$. That is the $d$-dimensional harmonic oscillator $V(r)=b^2 \, r^2$. We note that the determinant $\Delta_{n'+1}=0$ vanishes identically  only when $n'=2m,\,\, m=0,2,4,\dots$.  Further, using the  identity of the Biconfluent Heun function
\begin{equation}\label{eq56}
 H_B\left(k-2,0,4m+k,0,\sqrt{b}\, r\right)={}_1F_1\left(-m;k/2;b\,r^2\right),\qquad m=0,1,2,3,\dots,\,\, k=d+2l.
\end{equation}
that concides with the well-known solutions  of  the $d$-dimensional harmonic oscillator potential.
\subsection{Approximate solutions for arbitrary potential parameters}
\noindent For arbitrary values of the potential parameters $a$ and $b$ that do not necessarily obey the above conditions, we may use AIM directly to compute the eigenvalues \emph{accurately},  as the zeros of the termination condition (\ref{eq13}). The method can be used also to verify the exact solutions we have obtained in the above section. To use AIM, we start with
\begin{equation}\label{eq57}
\left\{ \begin{array}{l}
\lambda_0(r)=2br-\dfrac{k-1}{r} \\ \\
  s_0(r)= k\,b-E_{nl}^d+\dfrac{a}{r}.
       \end{array} \right.
\end{equation}
and computing the AIM sequences $\lambda_n$ and $s_n$  as given by Eq.(\ref{eq10}). We should note that for given values of the potential parameters $a$, $b,$ and of $k=d+2l$, the termination condition  $\delta_n=\lambda_n s_{n-1}-\lambda_{n-1}s_n=0
$ yields an expression that depends on both $r$ and $E$. In order to use AIM as an approximation technique for computing the eigenvalues $E$ we need to feed AIM with an initial value of $r=r_0$ that could stabilize AIM (that is, to avoid oscillations). For our calculations, we have found that $r_0=2$ stablizes AIM and allows us to compute the eigenvalues for arbitrary $k=d+2l$ and $n$ (the number of nodes) as shown in Table \ref{table:difdn3}. There is no magical assertion about $r_0=2$, indeed using an exact solvable case, say $E=5$ with $d=3,l=0, n'=1$ for $a=2$ and $b=1$, we may approximate $r=r_0$ by means of $E-V(r)=0$ which yields $r_0= 2$ as an initial starting value for the AIM process that can be increased slightly as for large $d$. The eigenvalue computations in Table \ref{table:difdn3}  were done using Maple version 13 running on an IBM architecture personal computer, where we used a high-precision environment. In order to accelerate our computation we have written our own code for a root-finding algorithm instead of using the default procedure {\tt Solve} of \emph{Maple 13}. The results of AIM may be obtained to any degree of precision, although we have reported our results for only the first eighteen decimal places, for example, we can easily find for $E_{nl}^d(a,b)$ that
\begin{align*}
E_{0,0}^{d=3}(a=1,b=1)&=~4.057~877~007~967~971~192~973~089~672~451~081~355~575~3_{N=66},\notag\\
E_{1,0}^{d=3}(a=1,b=1)&=~7.909~673~791~067~402~643~621~599~839~240~943~197~182~8_{N=65}.
\end{align*}

\begin{table}[h] \caption{Eigenvalues $E_{n0}^{d=2,3,4,5,6,7}$ for $V(r)=1/r+r^2$. The initial value utilize AIM is $r_0=2$. The subscript $N$ refer to the number of iteration used by AIM.\\ } 
\centering 
\begin{tabular}{|c|p{2.5in}||c|p{2.5in}|}
\hline
$n$&$E_{n0}^{d=2}$&$n$&$E_{n0}^{d=3}$\\ \hline
0&$~3.496~523~195~977~584~904_{N=68}$&0&$~4.057~877~007~967~971~193_{N=62}$\\
1&$~7.236~061~809~572~725~332_{N=62}$&1&$~7.909~673~791~067~402~644_{N=65}$\\
2&$11.087~207~289~903~431~629_{N=66}$&2&$11.819~201~619~422~902~597_{N=72}$\\
3&$14.987~686~167~769~085~806_{N=68}$&3& $15.755~974~584~087~041~187_{N=69}$ \\
4&$18.914~845~906~356~635~037_{N=70}$&4&$19.708~234~144~818~473~335_{N=72}$\\
5&$22.858~359~294~293~599~064_{N=76}$&5&$23.670~343~578~651~163~274_{N=80}$\\
6&$26.812~770~333~469~636~285_{N=77}$&6&$27.639~205~893~933~559~031_{N=75}$\\ \hline
$n$&$E_{n0}^{d=4}\equiv E_{n1}^{2}$&$n$&$E_{n0}^{d=5}\equiv E_{n1}^{3}$\\ \hline
0&$~4.855~342~290~384~481~116_{N=61}$&0&$~5.735~130~562~770~478~606_{N=62}$\\
1&$~8.759~375~855~335~329~641_{N=72}$&1&$~9.666~978~698~978~433~146_{N=65}$\\
2&$12.696~079~483~403~726~859_{N=66}$&2&$13.619~220~040~408~034~056_{N=67}$\\
3&$16.649~791~569~971~972~988_{N=69}$&3& $17.582~990~605~777~455~161_{N=77}$ \\
4&$20.613~775~425~537~580~344_{N=75}$&4&$21.554~094~076~075~464~896_{N=72}$\\
5&$24.584~567~825~802~389~703_{N=73}$&5&$25.530~235~189~253~178~567_{N=75}$\\
6&$28.560~170~841~418~040~482_{N=78}$&6&$29.510~030~701~250~179~290_{N=78}$\\
\hline
$n$&$E_{n0}^{d=6}\equiv E_{n1}^{4}\equiv E_{n2}^{2}$&$n$&$E_{n0}^{d=7}\equiv E_{n1}^{5}\equiv E_{n2}^{3}$\\ \hline
0&$~6.653~839~972~029~922~498_{N=64}$&0&$~7.594~350~931~424~006~160_{N=48}$\\
1&$10.602~367~239~032~036~476_{N=68}$&1&$11.553~756~993~287~284~639_{N=46}$\\
2&$14.564~582~581~426~144~447_{N=72}$&2&$15.522~859~985~850~837~447_{N=48}$\\
3&$18.535~063~827~411~992~187_{N=72}$&3&$19.498~137~514~526~264~446_{N=47}$ \\
4&$22.511~033~550~195~534~839_{N=72}$&4&$23.477~664~734~542~715~807_{N=46}$\\
5&$26.490~890~406~969~728~329_{N=75}$&5&$27.460~281~139~971~802~109_{N=45}$\\
6&$30.473~632~062~108~310~237_{N=47}$&6&$31.445~236~047~407~720~108_{N=45}$\\
\hline
\end{tabular}
\label{table:difdn3}
\end{table}

\section{Exact and approximate solutions for hard confinement}\label{spec}
\subsection{Analytic solutions}
\noindent We now consider the $d$-dimensional Schr\"odinger equation
\begin{equation}\label{eq58}
\left[-{d^2\over dr^2}+{(k-1)(k-3)\over 4r^2}+V(r)\right]u_{nl}^d(r)=E_{nl}^d u_{nl}^d(r),\quad 0<r< R,
\end{equation}
where, for $a>0$,
\begin{equation}\label{eq59}
V(r)=\left\{ \begin{array}{ll}
 {a\over r}+b^2r^2, &\mbox{ if $0<r< R$}, \\ \\
  \infty, &\mbox{ if $r\geq  R$,}
       \end{array} \right.
\end{equation}
and $u_{nl}^d(0)=u_{nl}^d(R)=0$. We may assume the following ansatz for the wave function
\begin{equation}\label{eq60}
u_{nl}^d(r)=r^{(k-1)/2}(R-r)\exp\left(-{b\over 2}~r^2\right)~f_n(r),\quad k=d+2l.
\end{equation}
where $R$ is the radius of confinement, and the $(R-r)$ factor ensures that the radial wavefunction $u_{nl}^d(r)$ vanishes at the boundary $r=R$. On substituting (\ref{eq60}) into (\ref{eq58}), we obtain the following second-order differential equation for the functions $f_n(r)$:
\begin{align}\label{eq61}
r\,(r-R)\,f_n''(r)&+[-2\,b\,r^3+2b\,R\,r^2+(k+1)r+(1-k)R]f_n'(r)\notag\\
&+{\bigg[(E_{nl}^d-b\,(k+2))r^2+(R(kb-E_{nl}^d)-a)r+Ra+k-1\bigg]}f_n(r)=0,\qquad b\neq 0
\end{align}
This differential equation cannot be studied using Theorem 2. However, the following theorem can be used.
\vskip0.1true in
\noindent{\bf Theorem 3 [\cite{saad2011} Theorem 3].} \emph{
The second-order linear differential equation \begin{align}\label{eq62}
(a_{4,0}r^4+a_{4,1}r^3+a_{4,2}r^2&+a_{4,3}r+a_{4,4})y^{\prime \prime}+(a_{3,0}r^3+a_{3,1}r^2+a_{3,2}r+a_{3,3})y'-(\tau_{2,0}r^2+\tau_{2,1} r+\tau_{2,2})y=0,
\end{align}
has a polynomial solution
$y(r)=\sum_{k=0}^nc_k r^k$
if
\begin{equation}\label{eq63}
\tau_{2,0}=n(n-1)~a_{4,0}+n~a_{3,0},\quad n=0,1,2,\dots,
\end{equation}
provided $a_{4,0}^2+a_{3,0}^2\neq 0$ where the polynomial coefficients $c_n$ satisfy the five-term recurrence relation
\begin{align}\label{eq64}
((n-2)&(n-3)a_{4,0}+(n-2) a_{3,0}-\tau_{2,0})c_{n-2}+((n-1)(n-2)a_{4,1}+ (n-1)a_{3,1}-\tau_{2,1})c_{n-1}\notag\\
&+(n(n-1)a_{4,2}+ na_{3,2}-\tau_{2,2})c_n+(n(n+1)a_{4,3}+(n+1) a_{3,3} )c_{n+1}+(n+2)(n+1)a_{4,4}c_{n+2}=0
\end{align}
with
$c_{-2}=c_{-1}=0$. In particular, for the zero-degree polynomials $c_0\neq 0$ and $c_n=0,~n\geq 1$, we must have $\tau_{2,0}=0$ along with
\begin{equation}\label{eq65}
\tau_{2,1}=0,~\tau_{2,2}=0.
\end{equation}
For the first-degree polynomial solution,  $c_0\neq 0,\quad c_1\neq 0$ and $c_n=0,n\geq 2$, we must have $
\tau_{2,0}=a_{3,0}
$
along with the vanishing of the two $2\times 2$-determinants, simultaneously,
\begin{equation}\label{eq66}
\left|\begin{array}{ccc}
-\tau_{2,2}&a_{3,3}\\
-\tau_{2,1}& a_{3,2} -\tau_{2,2}
\end{array}\right|=0,\quad
\mbox{and}\quad
\left|\begin{array}{ccc}
-\tau_{2,2}&a_{3,3}\\
-a_{3,0}& a_{3,1} -\tau_{2,1}
\end{array}\right|=0.
\end{equation}
For the second-degree polynomial solution, $c_0\neq0,c_1\neq0,c_2\neq0$ and $c_n=0$ for $n\geq 3$, we must have
$
\tau_{2,0}=2~a_{4,0}+2a_{3,0}$
along with the vanishing of the two $3\times 3$-determinants, simultaneously,
\begin{equation}\label{eq67}
\left|\begin{array}{ccc}
-\tau_{2,2}&a_{3,3}&2a_{4,4}\\
-\tau_{2,1}& a_{3,2} -\tau_{2,2}&2a_{4,3}+ 2a_{3,3}\\
-2a_{4,0}-2a_{3,0}&a_{3,1} -\tau_{2,1}&2a_{4,2}+2a_{3,2}-\tau_{2,2}
\end{array}\right|=0,~\mbox{and}~
\left|\begin{array}{ccc}
-\tau_{2,2}&a_{3,3}&2a_{4,4}\\
-\tau_{2,1}& a_{3,2} -\tau_{2,2}&2a_{4,3}+ 2a_{3,3}\\
0&-2a_{4,0}-a_{3,0}&2a_{4,1} +2a_{3,1}-\tau_{2,1}
\end{array}\right|=0,
\end{equation}
For the third-degree polynomial solutions $c_0\neq0,~c_1\neq0,~c_2\neq0,~ c_3\neq 0$ and $c_n=0$ for $n\geq 4$, we must have
$
\tau_{2,0}=6~a_{4,0}+3~a_{3,0}$ along with
 \begin{equation}\label{eq68}
\left|\begin{array}{cccc}
-\tau_{2,2}&a_{3,3}&2a_{4,4}&0\\
-\tau_{2,1}& a_{3,2} -\tau_{2,2}&2a_{4,3}+ 2a_{3,3}& 6a_{4,4}\\
-\tau_{2,0}&a_{3,1} -\tau_{2,1}&2a_{3,2}+2a_{4,2}-\tau_{2,2}&3a_{3,3}+6a_{4,3}\\
0&-2a_{3,0}-6a_{4,0}&2a_{3,2}+2a_{4,1}-\tau_{2,1}&3a_{3,2}+6a_{4,2}-\tau_{2,2}
\end{array}\right|=0,
\end{equation}
and
\begin{equation}\label{eq69}
\left|\begin{array}{cccc}
-\tau_{2,2}&a_{3,3}&2a_{4,4}&0\\
-\tau_{2,1}& a_{3,2} -\tau_{2,2}&2a_{4,3}+ 2a_{3,3}& 6a_{4,4}\\
-\tau_{2,0}&a_{3,1} -\tau_{2,1}&2a_{3,2}+2a_{4,2}-\tau_{2,2}&3a_{3,3}+6a_{4,3}\\
0&0&-a_{3,0}-4a_{4,0}&3a_{3,1}+6a_{4,1}-\tau_{2,1}
\end{array}\right|=0,
\end{equation}
and so on, for higher-order polynomial solutions. The vanishing of these determinants can be regarded as the conditions under which the coefficients $\tau_{2,1}$ and $\tau_{2,2}$ of Eq.(\ref{eq54}) are determined.
}
\vskip0.1true in
\noindent Using Theorem 3, we may note using $a_{4,0}=a_{4,1}=a_{4,4}=0,a_{4,2}=1,a_{4,3}=-R,a_{3,0}=-2\,b,a_{3,1}=2\,b\,R,a_{3,2}=(k+1),a_{3,3}=(1-k)R,\tau_{2,0}=-E_{nl}^d+b\,(k+2),\tau_{2,1}=-R(k\,b\,-E_{nl}^d)+a,\tau_{2,2}=-Ra-k+1$, that the necessary condition for polynomial solutions $f_n(r)=\sum_{k=0}^n c_k r^k$ of Eq.(\ref{eq53}) is
\begin{equation}\label{eq70}
E_{nl}^d=b\, (2n+k+2),\qquad k=d+2l,
\end{equation}
where $n'=n+1$ refers to the degree of the polynomial solution of equation \eqref{eq60}  and not necessarily equal to the number of nodes of the exact wave function. For $a>0$, it is important to note from Table \ref{table:nonlin2} that the exact solutions, for the hard confined case, exist only for $n'\geq 2$. Indeed, using the determinants of equation \eqref{eq69} for $n=2$, we have
\begin{align}\label{eq71}
\left\{ \begin{array}{l}
 6 k (k^2-1) +
  6 a k (k+1) R + (3 a^2 (k+1) -
     4b (k-1) (6+7k)) R^2 +
  a (a^2 + 4 b (2- 5k)) R^3 = 0,\\ \\
  4 b k R (aR-1 + k) - (a + 4 bR) \left(2 k(k-1 ) +
     2 a k R + (a^2 - 8 b (k-1)) R^2\right) = 0.
       \end{array} \right.
\end{align}
With $k=3 ~(i.e.~ d=3,~l=0),b=1$, the roots of the equations \eqref{eq71} gives
$$a=2.293~766~824~743~528~3\qquad \mbox{and}\qquad R=1.447~082~228~754~503,\quad\mbox{where}\quad E_{20}^3=9,$$
as we expected using \ref{table:nonlin2}.
 Further, for $n=4$, equations \eqref{eq68} and \eqref{eq69} implies
\begin{align}\label{eq72}
\left\{ \begin{array}{l}
 24 k (k^2-1) (2 + k) + 24 a k (1 + k) (2 + k)R +
 12 (a^2 - 8 b (k-1 )) (1 + k) (2 + k) R^2 +
 4 a (a^2 + 4 b (2 - 5 k)) (2 + k) R^3\\
 + (a^4 - 4 a^2 (b + 8 b k) +
    96 b^2 (-1 + k^2)) R^4 = 0,\\ \\
6 k (k^2-1 ) + 6 a k (1 + k) R +
 3 (a^2 - 8 b (-1 + k)) (1 + k) R^2 + a (a^2 + 4 b (2 - 5 k)) R^3 +
 2 b (a^2 + 4 b (2 - 5 k)) R^4= 0.
       \end{array} \right.
\end{align}
With $k=3 ~(i.e.~ d=3,~l=0),b=1$, the roots of the equations \eqref{eq71} gives
$$a=4.970~009~395~208~089\equiv\sqrt{70-6\sqrt{57}}\qquad \mbox{and}\qquad R=1.653~264~540~801~603~3,\quad\mbox{where}\quad E_{20}^3=11,$$
as we, again, expected using \ref{table:nonlin2}.  Similar exact results can be obtained for higher $n'$ (the degree of the polynomial solutions).
\subsection{Approximate solutions for hard confinement and arbitrary parameters}
\noindent For the arbitrary values of $a, b$ and $R$, not necessarily satisfying the  above special conditions, we may use AIM directly
to compute the eigenvalues with a very high degree of accuracy.  This also allows us to verify the exact solutions we obtained in the perevious sections. Similarly to the soft-confined case, we start the iteration of the AIM sequence $\lambda_n$ and $s_n$ with
\begin{equation}\label{eq73}
\left\{ \begin{array}{l}
\lambda_0(r)=\dfrac{1-k}{r} + 2\, b\, r - \dfrac{2}{(r - R)}, \\ \\
  s_0(r)=b\,(2 +k)-E+\dfrac{a\, R+k-1}{r\,R}+\dfrac{2\, b\,R^2-k+1}{R\,(r-R) },
      \end{array} \right.
\end{equation}
where $0<r<R$. We may note for $R\rightarrow \infty$, equation \eqref{eq73} reduces to \eqref{eq57}. It is interesting to note in this case, that, unlike the soft-confined case, the roots of the termination condition $\delta_n=\lambda_ns_{n-1}-\lambda_{n-1}s_n=0$ are much easier to handle in the present case. This is due to the fact that $r_0$ is now bound within $(0,R)$ for every given $R$. Thus, it is sufficient to start our iteration process with initial value $r_0=R/2\sim 0.5$. In table \ref{table:difdn4},
we report the eigenvalues we have computed using AIM for a fixed radius of confinement $R = 1.447~082~228~754~501~502$, with $a = \sqrt{30-6\sqrt{17}}$ and $b=1$ as an initial
value to seed the AIM process.

\begin{table}[h] \caption{Eigenvalues $E_{n0}^{d=3}(a,b;R)$ for $V (r) = a/r + br^2, r\in (0,R)$, where $a =\sqrt{30-6\sqrt{17}} $, $b = 1$, $R = 1.447~082~228~754~501~502$ and different of $n$. The subscript $N$ refers to the number of iteration used by AIM.\\ } 
\centering 
\begin{tabular}{|c|c|}
\hline
$n$&$E_{n0}^{d=3}(a =\sqrt{30-6\sqrt{17}},b=1;R = 1.447~082~228~754~501~502)$\\ \hline
0&$9.000~000~000~000~000~001_{N=5~(Exact)}$\\
\hline
1&$24.305~412~213~817~055~207_{N=35}$\\
\hline
2&$48.570~802~600~950~511~528_{N=34}$\\
\hline
3&$82.052~426~304~188~379~099_{N=41}$\\
\hline
4&$124.845~251~820~221~004~239_{N=50}$\\
\hline
5&$238.517~551~072~045~582~565_{N=64}$\\
\hline
\end{tabular}
\label{table:difdn4}
\end{table}

\noindent In general, the computation of the eigenvalues is fast, as is illustrated by the small
number of iteration $N$ in Tables \ref{table:difdn5}. The same procedure can be applied to compute the eigenvalues for other values of $a$, $b$ $R$, and arbitrary dimension $d$. The results of AIM may be obtained to any degree of precision, although we have reported our results for only the first eighteen decimal places. It is clear from the table that our results confirm the invariance of the eigenvalues under the transformation $(d,l)\rightarrow (d\mp2,l\pm 1).$
\begin{table}[h] \caption{Eigenvalues $E_{nl}^{d}(a,b;R)$ for $V (r) = a/r + br^2, r\in (0,R)$, where $a = 1$, $b =1$, $R = 1$ and different $n$ and $l$. The subscript $N$ refers to the number of iteration used by AIM.\\ } 
\centering 
\begin{tabular}{|l|l|l|}
\hline
$n$&$l$&$E_{nl}^{d=2}(1,1;1)$\\ \hline
0&0&$~~9.298~213~743~966~306~503_{N=29}$\\
~&1&$~17.056~214~768~511~049~448_{N=25}$\\
~&2&$~28.503~765~718~945~267~353_{N=25}$~$\equiv E_{01}^{d=4}$\\
~&3&$~42.737~355~022~574~771~999_{N=24}$~$\equiv E_{01}^{d=6}\equiv E_{02}^{d=4}$\\
~&4&$~59.564~456~107~008~915~084_{N=26}$~$\equiv E_{01}^{d=8}\equiv E_{02}^{d=6}\equiv E_{03}^{d=4}$\\
~&5&$~78.892~555~598~221~694~492_{N=29}$~$\equiv E_{01}^{d=10}\equiv E_{02}^{d=8}\equiv E_{03}^{d=6}\equiv E_{04}^{d=4}$\\
\hline\hline
$n$&$l$&$E_{nl}^{d=3}(1,1;1)$\\ \hline
0& 0&$~12.550~092~461~190~652~257_{N=26}$\\
&1&$~22.410~590~350~956~293~454_{N=25}$~~$\equiv E_{00}^{d=5}$\\
&2&$~35.288~239~785~280~558~264_{N=23}$~~$\equiv E_{00}^{d=7}\equiv E_{01}^{d=5}$\\
&3&$~50.833~639~418~866~620~375_{N=25}$~~$\equiv E_{00}^{d=9}\equiv E_{01}^{d=7}\equiv E_{02}^{d=5}$\\
&4&$~68.920~051~722~100~849~182_{N=28}$~~$\equiv E_{00}^{d=11}\equiv E_{01}^{d=9}\equiv E_{02}^{d=7}\equiv E_{03}^{d=5}$\\
&5&$~89.475~411~786~048~045~561_{N=30}$~~$\equiv E_{00}^{d=13}\equiv E_{01}^{d=11}\equiv E_{02}^{d=9}\equiv E_{03}^{d=7}\equiv E_{04}^{d=5}$\\
\hline\hline
$n$&$l$&$E_{nl}^{d=4}(1,1;1)$\\ \hline
0& 0&$~17.056~214~768~511~049~448_{N=25}$\\
&1&$~28.503~765~718~945~267~353_{N=25}$~~$\equiv E_{00}^{d=6}$\\
&2&$~42.737~355~022~574~771~999_{N=24}$~~$\equiv E_{00}^{d=8}\equiv E_{01}^{d=6}$\\
&3&$~59.564~456~107~008~915~084_{N=26}$~~$\equiv E_{00}^{d=10}\equiv E_{01}^{d=8}\equiv E_{02}^{d=6}$\\
&4&$~78.892~555~598~221~694~492_{N=29}$~~$\equiv E_{00}^{d=12}\equiv E_{01}^{d=10}\equiv E_{02}^{d=8}\equiv E_{03}^{d=6}$\\
&5&$~100.663~030~250~522~172~574_{N=32}$~~$\equiv E_{00}^{d=14}\equiv E_{01}^{d=12}\equiv E_{02}^{d=10}\equiv E_{03}^{d=8}\equiv E_{04}^{d=6}$\\
\hline
\end{tabular}
\label{table:difdn5}
\end{table}

\section{Conclusion}\label{conc}
We study a model atom-like system $-\half\Delta -a/r$ which is confined softly by the inclusion of a harmonic-oscillator potential term $b\,r^2$ and possibly also by the presence of an impenetrable spherical box of radius $R.$ For $b > 0$ or $R < \infty,$ the entire spectrum $E_{n,\ell}^d(a,b,R)$ is discrete.  We have studied these eigenvalues and we present an approximate spectral formula for the `free' case, $R = \infty$.  For the general case of $R\le\infty,$ AIM has been used to provide both a large number of exact analytical solutions, valid for certain special choices of the parameters $\{a,b,R\},$ and also very accurate numerical eigenvalues for arbitrary parametric data.
In the cases where we have found analytic solutions for $R=\infty$, the exact wave functions are no longer expressed in terms of known special functions, as is possible for the hydrogen atom.  However, the exact solutions we have found for confining potentials correspond to confinement at the zeros of the unconfined case. An interesting qualitative feature seems to be that $E_{n,\ell}^d(a,b,R)$, for large $R$, is concave with respect to $n$, $\ell$,
or $d$, but becomes convex as $R$ is reduced; this may arise because the reduction in $R$ perturbs the higher states more severely since, when free, they are naturally  more spread out. It is hoped that the work reported in the present paper will provide a useful addition to the growing body of results concerning the spectra of confined atomic systems in $d$ dimensions.

\section{Acknowledgments}
\medskip
\noindent Partial financial support of this work under Grant Nos. GP3438 and GP249507 from the
Natural Sciences and Engineering Research Council of Canada
 is gratefully acknowledged by two of us (RLH and NS). KDS thanks the Department of Science and Technology,
New Delhi, for the J.C. Bose fellowship award. NS and KDS are grateful for the
  hospitality provided by the Department of Mathematics and Statistics of
Concordia University, where part of this work was carried out.  RLH is grateful for the hospitality provided by the School of Chemistry, University  of Hyderabad 500046, India, where part of this work was carried out.


\begin{thebibliography}{00}


\bibitem{cruz} J. R. Sabin, E. J. Br\"{a}ndas and S. A. Cruz (Eds.), \emph{Advances in quantum chemistry: theory of confined quantum systems - Part one} (book \textbf{57}), Academic Press. New York, 2009.
\bibitem{sen} K. D. Sen (Ed.), \emph{Electronic structure of quantum confined atoms and molecules},  Springer (UK) 2014.
\bibitem{sajeev}Y. Sajeev and N. Moiseyev, Phys. Rev. B {\bf 78}, 075316 (2008).
\bibitem{genkin}M. Genkin  and E. Lindroth, Phys. Rev. B {\bf 81}, 125315 (2010).
\bibitem{chak}S. Chakraborty and Y. K. Ho, Phys. Rev. A  {\bf 84}, 032515 (2011).
\bibitem{taut} M. Taut, Phys. Rev. A {\textbf 48}, 3561 (1993).
\bibitem{pupyshev98} A.~I.~Pupyshev and A.~V.~Scherbinin, Chem.~Phys.~Lett.~ \textbf{295}, 217 (1998); Phys.~Lett.~ A \textbf{299}, 371 (2002).
\bibitem{Herrick} D. R. Herrick , J. Math. Phys. \textbf{16}, 281 (1975).
\bibitem{HS} D. R. Herrick and F. H. Stillinger, Phys. Rev. A \textbf{11}, 42 (1975).
\bibitem{senkd} K.D. Sen, H.E. Montgomery, Jr. and N.A.Aquino, Int. J. Quantum Chem. \textbf{107}, 798 (2007).
\bibitem{spm} K.D. Sen, V. I. Pupyshev and H.E. Montgomery Jr., Ad. Quantum Chem. \textbf{57}, 25 (2009).
\bibitem{Jaber} Muzaian A. Shaqqor and Sami M. AL-Jaber, Int. J. Theor. Phys. \textbf{48},
2462 (2009).
\bibitem{Ed} H.E. Montgomery Jr , G. Campoy and N. Aquino, Phys.
Scr. \textbf{81}, 045010 (2010).
\bibitem{gusun}Xiao-Yan Gu and Jian-Qiang Sun , J. Math. Phys. {\bf 51}, 022106 (2010). 
\bibitem{agboola}D. Agboola , Pramana {\bf 76},  875 (2011). 
\bibitem{louck} J. D. Louck,  J. Mol. Spectrosc. 4 (1960) 298-333; A. Chatterjee,
Phys. Rep. 186 (1990) 249-370.
\bibitem{atkin}K. Atkinson and W. Han,{\it Spherical Harmonics and Approximations on the Unit Sphere: An Introduction}
(Springer, New York, 2012).
\bibitem{Doren} D. J. Doren and D. R. Herschbach,  J. Chem. Phys. 85 (1986) 4557.
\bibitem{Fock} V. A. Fock, Bull. Acad. Sci USSR, Phys.
Ser., \textbf{2}, 169 (1935).
\bibitem{alliluev58} S.~P.~Alliluev, Sov.~Phys.~JETP \textbf{6}, 156 (1958).
\bibitem{Avery} J. Avery, {\it Hyperspherical Harmonics: Applications in Quantum Theory}
,Kluwer Academic, Boston (1989).
\bibitem{Avery2} D. R. Herschbach, J. Avery, and O. Goscinski (Eds.), Dimensional Scaling in Chemical Physics,
Kluwer Academic, Dordrecht (1993).
\bibitem{Singer} S. F. Singer, \emph{Linearity, Symmetry, and Prediction in the Hydrogen Atom}, (Springer, New York, 2005). [$SO(4)$-symmetry of the Hydrogen atom is discussed in Chapters 8 and 9.]
\bibitem{Fradkin} D. M. Fradkin, Amer. J. Phys. \textbf{33}, 207 (1965).

\bibitem{Chatt} A. Chatterjee, Phys. Rep. \textbf{186}, 249 (1990).
\bibitem{Dunn} M.Dunn and D.K. Watson, Ann. Phys. \textbf{251}, 266
(1996).
\bibitem{Ma} Xiao-Yan Gu and Zhong-Qi Ma, J. Math. Phys. 44,3763
(2003).
\bibitem{Night} M.P. Nightingale and Mervlyn Moodley, J. Chem.
Phys. 123, 014304 (2005).
\bibitem{Shannon} C.~E. Shannon, A mathematical theory of communication, The Bell
System Technical Journal \textbf{27}, 379--423 (1948) $;$ ibid,
\textbf{27}, 623 (1948).
\bibitem{BBM} I. Bialynicki-Birula and J.
Mycielski, Commun. Math. Phys. \textbf{44}, 129 (1975).
\bibitem{Fisher}
R.~A. Fisher, Theory of statistical estimation, in: Proceedings of
the Cambridge Philosophical Society, no.~22,  pp. 700--725 (1925).
\bibitem{Dehesa} Elvira Romera, P. S\'anchez-Moreno, and J. S. Dehesa, J. Math. Phys. 47, 103504 (2006)
, Mol. Phys. 108, 2527 (2010).


\bibitem{env1}R.~L.~Hall, Phys. Rev. D {\bf 22},2062 (1980).
\bibitem{env2}R.~L.~Hall, J. Math. Phys. {\bf 24}, 324 (1983).
\bibitem{env3}R.~L.~Hall, J. Math. Phys. {\bf 25}, 2708 (1984).
\bibitem{env4}R.~L.~Hall, Phys. Rev. A {\bf 39}, 550 (1989).
\bibitem{env5}R.~L.~Hall, J. Math. Phys. {\bf 33}, 1710 (1992).
\bibitem{env6}R.~L.~Hall, J. Math. Phys. {\bf 34}, 2779 (1993).

\bibitem{GS}S.~J.~Gustafson and I.~M.~Sigal, {\it Mathematical concepts of
quantum mechanics}, (Springer, New York, 2006). [The operator
inequality is proved for dimensions $d\ge 3$ on page 32.]
\bibitem{RS2}M.~Reed and B.~Simon, {\it Methods of modern mathematical physics II:
Fourier analysis and self-adjointness}, (Academic Press, New york,
1975). [The operator inequality is proved for $d=3$ on p 169].
\bibitem{common}A.~K.~Common, J. Phys. A {\bf 18}, 2219 (1985).
\bibitem{thirring}W. Thirring, {\it A Course in Mathematical Physics 3: Quantum Mechanics of Atoms and Molecules}, (Springer, New York/Wien, 1990). The local energy theorem  is discussed on p154.
\bibitem{aim} H. Ciftci, R. L Hall and N. Saad,  J. Phys. A: Math. Gen. 36 (2003) 11807.
\bibitem{aim1} N. Saad, R. L. Hall, and H. Ciftci,  J. Phys. A: Math. Gen. 39 (2006) 13445-13454.
\bibitem{saad} H. Ciftci, R. L. Hall, N. Saad, and E. Dogu,  J. Phys. A: Math. Theor. 43 (2010) 415206.
\bibitem{saad2011} R. L.. Hall, N. Saad, and K. Sen, \emph{Discrete spectra for confined and unconfined $-a/r + b r^2$ potentials in $d$-dimensions},  	J. Math. Phys. 52 (2011) 092103
\bibitem{landau} L. D. Landau and E. M. Lifshitz, \emph{Quantum Mechanics: non-relativistic theory,} Pergamon, London, 1981.
\bibitem{arscott} F. M. Arscott, \emph{Periodic Differential Equations: An Introduction to Mathieu, Lam\'e, and Allied Functions,} Pergamon Press (1964).
\bibitem{hautot1969} A. Hautot, \emph{Sur les solutions polynomiales de l'equation differentielle de Heun}, Bull. Soc. Roy. Sci. Li\'ege 38 (1969) 654 - 659 and 660 - 663.
\bibitem{rovder} J. Rovder, \emph{Zeros of the polynomial solutions of the differential equation $xy'' +(\beta_0+\beta_1\,x+\beta_2\, x^2)y'+(\gamma-n\beta_2x)y=0$,} Mat. C\u{a}s. \textbf{24} (1974) 15 - 20..
\end{thebibliography}
\end{document}